\documentclass[11pt]{article}
\usepackage{epsfig}
\def\be{\begin{eqnarray}}\def\ba{\begin{eqnarray}}
\def\ee{\end{eqnarray}}\def\ea{\end{eqnarray}}
\def\ben{\begin{enumerate}}\def\bitem{\begin{itemize}}
\def\een{\end{enumerate}}\def\eitem{\end{itemize}}
\def\no{\nonumber\\}

\newcommand{\e}{{\mbox{e}}}

\def\roughly#1{\mathrel{\raise.3ex\hbox{$#1$\kern-.75em%
\lower1ex\hbox{$\sim$}}}}

\def\A0{A_0}
\def\bq{\begin{equation}}
\def\eq{\end{equation}}
\def\la{\langle}\def\ra{\rangle}
\def\K0{K^0}

\begin{document}
\begin{titlepage}
\begin{center}

 \vskip 1.5cm

{\Large \bf Heavy quarkonium in a holographic QCD model } \vskip
1. cm
  {\large Youngman Kim$^{(a)}$, Jong-Phil Lee$^{(a)}$, Su Houng
    Lee$^{(b)}$ }

\vskip 0.5cm
(a)~{\it School of Physics, Korea Institute for Advanced Study , Seoul
  130-722, Korea}

(b)~{\it Institute of Physics and Applied Physics,
Yonsei University, Seoul 120-749, Korea}

\end{center}

\centerline{(\today) }
\vskip 1cm
\vspace{1.0cm plus 0.5cm minus 0.5cm}

\begin{abstract}

Encouraged by recent developments in AdS/QCD models for light
quark system, we study heavy quarkonium in the framework of the
AdS/QCD models.  We calculate the masses of $c\bar c$ vector meson
states using the AdS/QCD models at zero and at finite temperature.
Among the models adopted in this work, we find that the soft wall
model describes the low-lying heavy quark meson states at zero
temperature relatively well.   At finite temperature, we observe
that once the bound state is above $T_c$, its mass will increase
with temperature until it dissociates at a temperature of around
$494~{\rm MeV}$. It is shown that the dissociation temperature is
fixed by the infrared cutoff of the models. The present model
serves as a unified non perturbative model to investigate the
properties of bound quarkonium states above $T_c$.

\end{abstract}

\end{titlepage}

\newpage

\section{Introduction}
Recent developments in AdS/CFT~\cite{adscft} find many interesting
possibilities to study strongly interacting system such as QCD.
Confinement is assured with an IR cut off in the
 AdS space~\cite{polchinski}, and
flavors are introduced by adding extra probe branes~\cite{karch}.
Phenomenological models were also suggested
  to construct a holographic model dual to
  QCD~\cite{EKSS,PR,Brodsky}.
 However, the meson spectrum of the models does not
follow the well-known Regge trajectories.
In Ref.~\cite{KKSS}, a phenomenological dilaton background was introduced
to improve the meson spectrum.  Remarkably such a
 dilaton-induced potential gives exactly the linear trajectory
of the meson spectrum: $m_n^2\sim n$. The model with a
phenomenological dilaton is further  improved by a finite UV
cutoff~\cite{ET}.

In this work, we delve into the spectrum of heavy quarkonium
states based on the AdS/QCD models~\cite{EKSS,PR}.   The
properties of heavy quark system both at zero and at finite
temperature have been the subject of intense investigation for
many years (for a review see, for example,~\cite{hQnR}).
 This is so because, at zero temperature, the
charmonium spectrum reflects detailed information about
confinement and interquark potentials in QCD\cite{potential}, and
at finite temperature, its change and dissociation will convey
signals of the QCD deconfining phase transition from a
relativistic heavy ion collisions (RHIC)\cite{MS86}.   In
addition, recent lattice calculations suggest that the charmonium
states will remain bound at finite temperature to about 1.6 to 2
times the critical temperature $T_c$\cite{AH04,Datta03}. This
suggests that analyzing the charmonium data from heavy ion
collision inevitably requires more detailed information about the
properties of charmonium states in QGP, such as the
 effective dissociation cross section and its dependence on the
charmonium velocity\cite{PBM,SHL}.  Therefore, it is very
important and a theoretical challenge to develop a consistent non
perturbative QCD picture for the heavy quark system both below and
above the phase transition temperature.  As a first step towards
accomplishing these goals, we focus on the spectrum of heavy
vector meson at zero and at finite temperature. This is partly
 because, as it is well-known
in the AdS/QCD models for light quarks~\cite{EKSS,PR}, the bulk vector
field does not couple to the bulk scalar field, and so the vector
sector of the models is relatively simple to analyze.
We calculate the
mass spectrum of the vector meson in three different models at
zero temperature, and find that among the models considered, the
soft wall model best fits the experimental value.
 We then introduce the black hole background on the AdS space and
obtain the temperature dependent mass spectrum.

 It should be
stressed that we are primarily interested in the spectrum and
properties of the heavy quark system in the quark-gluon plasma
(QGP), {\it i.e}, above $T_c$. Therefore, while a recent
analysis~\cite{Herzog} of a Hawking-Page type transition\cite{HP}
in the AdS/QCD
 models claimed that the AdS black hole is unstable
at low temperature, roughly $T\le 200 {\rm MeV}$,  the use of the
AdS black hole background at higher temperature, such as in the
present work, is well justified.    In general, using the AdS
black hole background will not give any temperature dependence at
low temperature, since the thermal AdS metric is just the AdS
background with compactified Euclidean time.  Moreover, according
to the Hawking-Page transition, the AdS black hole is unstable in
the confined phase below $T_c$, so that we  may not be able to use
the black hole background to study the temperature dependence of
light quark systems in AdS/QCD models.  This means, that in the
light of the Hawking-Page transition, the AdS/QCD models for light
quarks may not be of much use, because those light states are
expected to dissolve above the phase transition, and the only
interesting temperature dependence is expected to occur below the
phase transition temperature.   As it is, for the light quark
system, the AdS/QCD models are about the mesons and the baryons at
zero temperature.  However, given the present interest in the
properties of the heavy quark system and the validity of AdS black
hole approach at high temperature, it is interesting to study the
temperature dependence of some physical quantities in AdS/QCD
models.  In this respect  heavy quark system at finite
temperature, especially above $T_c$, may be a good theme for the
AdS/QCD models.
\section{Mass spectrum in AdS/QCD models}
%
In the AdS/QCD models~\cite{EKSS,PR}, mesons made of  light
quarks, such as the  pions and $\rho$-mesons, are described by
gauged SU$(N_f)_L$$\times$SU$(N_f)_R$ chiral symmetry in AdS$_5$,
where the metric is given by \ba ds_5^2=\frac{1}{z^2}\biggl(
\eta_{\mu\nu}dx^\mu dx^\nu -dz^2 \biggr)\, . \ea In
Ref.~\cite{EKSS}, gauged SU$(2)_L$$\times$SU$(2)_R$ chiral
symmetry is adpoted to study the mesons such as pions and
rho-mesons, while with  gauged SU$(3)_L$$\times$SU$(3)_R$ chiral
symmetry~\cite{PR} one could investigate the properties of mesons
including strangeness.  In the present study, we will attempt to
generalize the approach to the charm sector and focus on the
charmonium mass spectrum.  Since the charm quark is much heavier
than those in the light sectors and the symmetry is mainly broken
by the explicit charm quark mass, we  will adopt a separate
U(1)$_L$$\times$U(1)$_R$ symmetry in AdS$_5$. In AdS$_5$, the
chiral symmetry breaking, both explicit and spontaneous, is
realized through a bulk scalar field profile near the boundary
$z\rightarrow 0$, according to an AdS/CFT dictionary~\cite{KW99}.
For instance, in ~\cite{EKSS,PR} the bulk scalar field,
$\phi(x,z)$, which couples to the quark bilinear operator of $\bar
q q$, takes the following form near the boundary: \ba \phi\sim c_1
z+ c_2z^3\, , \ea where $c_1=m_q$ and $c_2\sim\la \bar qq \ra$. In
the light quark system, we could take the chiral-limit, $m_q=0$.
In the heavy quark system, we expect to have
 \ba
\Phi\sim a_1 z+ a_2 z^3\, ,
\ea
where $a_1=m_Q$ and $a_2\sim\la \bar QQ \ra$.
Here $Q$ is for heavy
quarks. In the present work we consider charm quark only and study
the spectrum of $c\bar c$ at zero and at finite temperature. We
describe the heavy quarkonium in AdS$_5$ with gauged
U(1)$_L$$\times$U(1)$_R$ symmetry, which is explicitly broken by
the $m_Q$-term in the bulk field profile near the boundary.  Note
that heavy quark and light quark systems have different IR
cutoffs, as will be discussed below.  In the potential model, the
$1^-$ $J/\psi (3097)$ and $0^-$ $\eta_c(2980)$ can be described as
s-wave charm quarks in the spin triplet and singlet states
respectively, while the $1^+$ $\chi_{c1}(3510)$ and $0^+$
$\chi_{c0}(3415)$ with charm quarks in the p-wave. However, the
present picture of treating the spectrum of heavy quarkonium in
terms of explicit symmetry breaking is phenomenologically also
acceptable, because the lowest parity partners have non degenerate
mass, $i.e.$ in the spin 1 channel $m(\chi_{c1})-m(J/\psi)=$413
MeV, and in the spin 0 $m(\chi_{c0})-m(\eta_c)=$435 MeV.

\subsection{Hard-wall model}
First we consider the hard-wall model, whose
action is given by
\ba &&{\rm S}_{\rm HW} =\int d^4x dz
\sqrt{g}{\cal L}_5\, ,\no &&\,\,\,\, {\cal L}_5=
-\frac{1}{4g_5^2}(L_{MN}L^{MN} +R_{MN}R^{MN}
)+|D_M\Phi|^2+3|\Phi|^2\, ,\label{hQCD-HW}
\ea
where $D_M\Phi =\partial_M\Phi -iL_M\Phi +i\Phi R_M$, $g_5$ is a 5D
gauge coupling constant, $L_{MN}=\partial_M L_N-\partial_NL_M-i[L_M,L_N]$
 and
$L_M=L_M T_0$
with $T_0$ being a U(1) generator. In the present work, we follow the
convention in~\cite{EKSS}.
 The scalar field is defined by
$\Phi=S\e^{i\pi_0T_0}$ and $\langle S\rangle\equiv\frac{1}{2}v(z)$, where $S$
is a real scalar and $\pi_0$ is a pseudoscalar. In this model, the
5D-AdS space is compactified such that $z_0<z<z_m^H$, where
$z_0\rightarrow 0$ and $z_m^H$ is an infrared (IR) cutoff. In the
light quark system~\cite{EKSS, PR}, the value of $z_m^L$
 is fixed by
the rho-meson mass ($m_\rho$) at zero temperature:
 $m_\rho (\simeq 770~{\rm MeV})\simeq3\pi/(4z_m^L)$$\rightarrow$
 $1/z_m^L\simeq 320~{\rm
MeV}$.  Note that the heavy quarkonium is blind to $z_m^L$, since
for heavy quarkonium $z_0<z<z_m^H$ and $z_m^H<z_m^L$.
 The vector meson mass spectrum from (\ref{hQCD-HW}) is
given by~\cite{PR} \ba m_n\simeq
(n-\frac{1}{4})\frac{\pi}{z_m^H}\label{mVh}\, .
\ea
Now we use
this formula to calculate  vector $c\bar c$ mass spectrum. We use
the lowest  $J/\psi$ mass of $3.096~{\rm GeV}$, as an input to fix
$z_m^H$. Then we obtain $1/z_m^H \simeq 1.315 {\rm GeV}$, which is
close to the
 value of c-quark mass.  With this, the mass of
 the second resonance is predict to be  $\sim 7.224~{\rm GeV}$, which
is quite different from the known mass of $\psi'$ of  $ 3.686
~{\rm GeV}$.

\subsection{Soft-wall model}
In the soft wall model~\cite{KKSS}, dilaton background was
introduced for the Regge behavior of the spectrum, and we work
mostly in this framework. \ba {\rm S}_{\rm SW} =\int d^4x
dz\e^{-\Phi}{\cal L}_5  \, ,\label{hQCD-SW} \ea where $\Phi=cz^2$.
Here the role of the hard-wall IR cutoff $z_m^H$ is replaced by
dilaton-induced potential, and the potential is given by
$V(z)\simeq cz^2$ at IR.
The equation of motion for the vector is~\cite{KKSS}
\ba
\partial_z \left ( \e^{-B} \partial_z v_n \right )+ m_n^2 \e^{-B}v_n=0\, ,
\ea
where $B=cz^2+\log z$ and  $m_n^2$ is the 4D mass.
After the following change, $v_n=\e^{B/2} V_n$, we obtain an exactly
solvable Schr\"odinger type equation
\ba
-\partial_z^2 V_n +{\cal V}(z)V_n=m_n^2V_n\, ,
\ea
where ${\cal V}(z)=cz^2+3/(4z^2)$.
 Then the mass spectrum of vector field in this model is given by~\cite{KKSS}
\ba m_n^2=4(n+1)c\, ,
\ea
where $\sqrt{c}\sim 1/z_m^H$.
Again the
lowest mode ($n=0$) is used to fix $c$, $\sqrt{c}\simeq 1.55~{\rm
GeV}$. Then the mass of the second resonance is $m_1\simeq
4.38~{\rm GeV}$, which is ~$20\%$ away from the experimental value of
$3.686 ~{\rm GeV}$, and the third one  $m_3\sim 5.36~{\rm GeV}$.
\subsection{A braneless set-up}
In~\cite{CR}, the IR cutoff is naturally replaced by the gluon
condensate, in contrast to the hard-wall model, where the IR
cutoff is introduced by
 hand.
The background found in
Ref.~\cite{CR} is
\ba
ds^2=(\frac{R}{z})^2 \left
  (\sqrt{1-(\frac{z}{z_c})^8}\eta_{\mu\nu}dx^\mu dx^\nu
-dz^2   \right )\, .\label{bG1} \ea The equation of motion reads
\ba \left ( \partial_z^2 -\frac{1}{z}(\frac{4}{f_c}-3)\partial_z
+\frac{m_c^2}{\sqrt{f_c}} \right )V_\mu=0\, ,\label{EoMV1} \ea where
$f_c=1-(z/z_c)^8$ and $z_c^{-1}\sim m_c$. The value of $z_c$ is
fixed, again, by the lowest $c\bar c$ mass, and is given by
$z_c^{-1}= 1.29~{\rm GeV}$. The mass of the second resonance is
about $7 {\rm GeV}$.\\

Judging from the observed vector $c\bar c$ meson spectrum, we
conclude that the soft wall model  fits the low-lying charmonium
masses better than the other models discussed in the present paper.

\section{Finite temperature}
To obtain the temperature dependence of the mass spectrum, we work on
 5D AdS-Schwarzchild background,
which describes the physics of the finite temperature in
dual 4D field theory,
\ba
ds^2=\frac{1}{z^2}\biggl(f(z) dt^2-(dx^i)^2 -\frac{1}{f(z)}dz^2
\biggr),~~f(z)=1-(\frac{z}{z_h})^4,\label{BH}
\ea
where $i=1,2,3$. The Hawking temperature is given by $T=1/(\pi z_h)$.
A study~\cite{Herzog} of a Hawking-Page type transition in the AdS/QCD
 models claimed that the AdS black hole is unstable
at low temperature, roughly below $T_c\sim 200 ~{\rm MeV}$. In the
analysis, the contribution from mesons are not considered, since
they are suppressed
 by $1/N_c$ compared to the gravitational part, and consequently, the estimated
 critical temperature could change slightly.
 Nevertheless, if we respect
 the observation made in~\cite{Herzog}, the background
 in Eq.~(\ref{BH}) may not be relevant for describing the low-temperature
 regime, below $T_c$.
In the present work, however, we are mainly  interested in the heavy  quarkonium
 spectrum in the QGP, and so we will evaluate the temperature dependence of
$c\bar c$  states above the critical temperature of the Hawking-Page type
 transition calculated in~\cite{Herzog}. In this respect, the AdS/QCD
 models, originally developed for light mesons like pions, may find
 their relevance in describing heavy quark system at finite temperature.

The equation of motion for the vector field at finite temperature in
 the
soft-wall model is given by \ba [\partial_z^2
-\bigl(2cz+\frac{4-3f}{zf}\bigr )\partial_z +\frac{m^2}{f^2}
]V_i=0\, .\label{VmT}
\ea
If we take $c=0$ and introduce the IR-cutoff $z_m^H$, Eq.
(\ref{VmT}) is reduced to the equation of motion in the hard wall
model. Here we set $\partial_x V_i=0$, which corresponds to taking
$\vec
 q=0$, spatial momentum to be zero,
to define the mass at finite temperature in field theory.
We first calculate the temperature dependent mass in the hard wall
model, though it is not very successful in heavy quark system at
zero temperature. We find that the heavy  quarkonium masses
decrease with temperature, similar to the behavior observed in the
light meson system~\cite{GY}.  An interesting feature of the hard
wall model at finite temperature for the heavy quark system is
that we can roughly estimate the dissociation temperature in a
simple and clear way.  As it is nicely described in ~\cite{GY},
when $z_h<z_m^L$, the cutoff $z_m^L$ is no longer an IR cutoff of
the system, since $0<z<z_h$, and the system is in a deconfined
phase. In the light quark system, the critical temperature of the
deconfinement is given by $T_c=1/(\pi z_m^L)\sim 102 ~{\rm MeV}$.
\footnote{ We note here that an analysis based on a Hawking-Page
type
  transition~\cite{Herzog} in the hard wall model predicts
$T_c=2^{1/4}/(\pi z_m^L)\simeq 1.189/(\pi z_m^L)$. }
On the analogy of the light meson system, we claim that a critical
temperature defined by  $T_d^H=1/(\pi z_m^H)$($\sim 418~{\rm
MeV}$) can be identified as the dissociation temperature for the
heavy quark system.   The dissociation temperature of the heavy
quark system is not the critical temperature of phase transition,
but the point above which the bound heavy quark system in the
quark gluon plasma dissociates into open heavy quarks. From
lattice calculation of pure gauge theory, this dissociation
 is found to be around 1.6$T_c$\cite{AH04,Datta03},
where for pure gauge theory, $T_c \simeq 260$ MeV such that
1.6$T_c=416$ MeV. Therefore the predicted dissociation temperature
from the hard wall model $T_d^H\sim 418~{\rm MeV}$ is close to the
lattice result.

Now we solve Eq. (\ref{VmT}) in the soft wall model.
To this end, we have to supply two
boundary conditions. At UV, $z=0$, we impose $V_i(z=\epsilon)=0$
with $\epsilon\rightarrow 0$. The boundary condition at IR calls
for more caution. At zero temperature, as in ~\cite{KKSS}, the IR
boundary condition (at $z=\infty$) is given by the requirement
that the action should be finite.
Now we require, at finite temperature, again that the action should be
finite. Note that at finite temperature $z_h$ plays a role of the IR
cutoff.  If the integrand of the action behaves, however, as $\sim
1/(z-z_h)^n$, $n\ge 1$ near $z_h$, then the action will not be finite.
The relevant part of the action  near $z_h$ takes the following form:
\ba
S_{SW}\sim \int_\epsilon^{z_h}(z_h-z) (\partial_z V_i)^2 dz\, .
\ea
Near $z_h$, the vector field $V_i$ will have the following profile:
 $V_i(z)\sim (z_h-z)^a$. To make the action finite near $z_h$, we
 require that $a>1/2$. And so we impose the IR boundary at $z_h$
as  $V_i(z=z_h)=0$.
\footnote{We refer to~\cite{KSJL} for some discussion on the IR
boundary condition in the soft wall model at finite temperature.
It is shown that in some case,  zero frequency and
zero momentum limit, the infalling boundary condition can be effectively
described by a
 Dirichlet boundary condition. }
\begin{figure}
\centerline{\epsfig{file=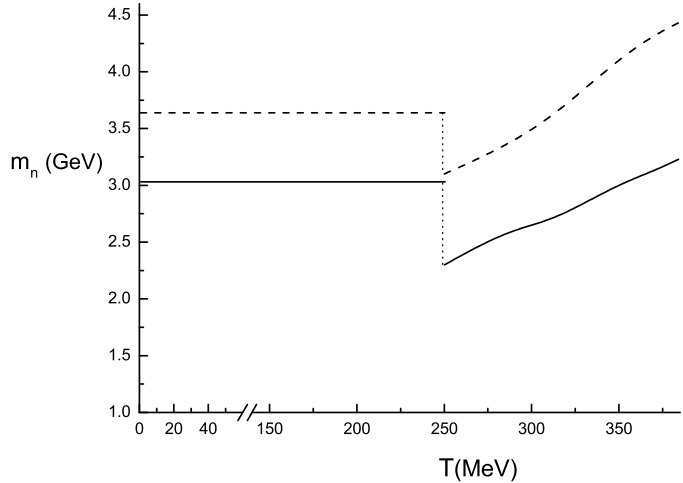,width=10.0cm}} \caption{\small
The mass of $c\bar c$ bound state in the vector channel at
  finite temperature, obtained in the soft wall model.  Here we show
  the first two modes, $n=0,1$.}
\label{cc2H}
\end{figure}
The results of the numerical solutions are shown in
Fig.~\ref{cc2H}, where we used the thermal AdS metric below $T_c$
and the AdS black hole above $T_c$. Here we set $T_c\simeq
250~{\rm MeV}$. As it is well-known, we have no
temperature dependence with the thermal AdS background, and so the
mass is temperature independent below $T_c$. As in
Fig.~\ref{cc2H}, we find that the mass will increase with
temperature above $T_c$. The sudden change of the mass spectrum
near $T_c$ is due to the use of different background below and
above $T_c$, namely, the thermal AdS and AdS black hole
respectively. Finally we estimate the dissociation temperature of
the heavy  quarkonium system in the soft wall model. Since $\sqrt{c}$
corresponds to $1/z_m^H$ in the hard wall model, we may define
$T_d^H=\sqrt{c}/\pi$, and so the predicted dissociation
temperature in the soft wall model is $\sim 494~{\rm MeV}$.

All the features appearing in Fig~\ref{cc2H}, namely the sudden
decrease at $T_c$ and its subsequent increase in the charmonium
mass, can be qualitatively understood using lattice results on the
color singlet part of the potential between a heavy quark and an
anti-quark\cite{Karsch01,Karsch03}.  The Lattice calculations of
the heavy quark potential show a sudden flattening of the
asymptotic value of the potential at $T_c$.   This could be
interpreted as the disappearance of the confining part of the
potential, namely the string tension at the phase transition,
which is first order for QCD with no dynamical quarks. Since the
$J/\psi$ has a non trivial part of the wave function resting on
the confining part, once this part vanishes, the mass is expected
to decrease suddenly, as long as it is still bound. Such a sudden
decrease of the $J/\psi$ mass just across the phase transition
boundary is also found in a recent QCD sum rule analysis with
condensate inputs from lattice gauge theory\cite{ML07}.
Furthermore, as the temperature increases further above $T_c$, the
potential becomes shallow and the binding energy is expected to
decrease\cite{Wong04}.  Hence the mass of the bound state will
increase until it dissociates into open charms with bare and
thermal masses.

\section{Summary}

We have applied three  AdS/QCD models to investigate properties of
heavy quark system at zero and at finite temperature above $T_c$.
We find that the soft wall model approaches  the low-lying heavy
quarkonium mass states at zero temperature better than the other two
models considered in this work.   At finite temperature, we
observe that above $T_c$ the masses of quarkonium states
 increase with
temperature, and the dissociation temperature, which is determined by
the IR cutoff of a  model, is around $494~{\rm
MeV}$ in the soft wall model.
 While we have limited the study to the calculation of mass
spectrum so far, the results seem encouraging and consistent with
expectations from lattice gauge theory calculation.

\vskip 1cm \noindent {\large\bf Acknowledgments}\\ The work of SHL
has been supported  by the Korea Research Foundation
KRF-2006-C00011.

\end{document}